\documentclass[twocolumn,showpacs,preprintnumbers,amsmath,amssymb]{revtex4}
\usepackage{graphicx}
\usepackage{dcolumn}
\usepackage{bm}
\begin{document}
\preprint{xxx}
\title{A Kelvin-wave cascade on a vortex in superfluid $^4$He at a very low
temperature}
\author{W.F. Vinen}
\affiliation{School of Physics and Astronomy, University of Birmingham,
\\Birmingham B15 2TT, United Kingdom}
\author{Makoto Tsubota and Akira Mitani}
\affiliation{Department of Physics, Osaka City University, \\Sumiyoshi-ku,
Osaka 558-8585,  Japan}
\date{\today}

\begin{abstract}
A study by computer simulation is reported of the behaviour of a quantized
vortex line at a very low temperature when there is continuous excitation of
low-frequency Kelvin waves.  There is no dissipation except by phonon
radiation at a very high frequency.  It is shown that non-linear coupling
leads to a net flow of energy to higher wavenumbers and  to the development
of a simple spectrum of Kelvin waves that is insensitive to the strength and
frequency of the exciting drive. The results are likely to be relevant to
the decay of turbulence in superfluid $^4$He at very low temperatures.
\end{abstract}
\maketitle

It is well known that quantized vortices can be formed in superfluid $^4$He.
Such vortices can support a transverse and circularly polarized wave motion
(a Kelvin wave),  with the approximate dispersion relation for a rectilinear
vortex \cite{donnelly1}
%Eq 1
\begin{equation}   \omega = \frac{\kappa
k^2}{4\pi}\bigg[\ln\Big(\frac{1}{ka}\Big)+c\bigg],
\end{equation}
where $\kappa$ is the quantum of circulation ($h/m_{4}$), $a$ is the vortex
core parameter, and $c\sim 1$.   The existence of these waves in an inviscid
fluid was first discussed as a theoretical possibility in the nineteenth
century \cite{thomson1},  but an experimental study in a fluid without
viscosity had to await the discovery of quantized vortices in superfluid
$^4$He.  Kelvin waves in uniformly rotating superfluid $^4$He were first
observed experimentally by Hall \cite{hall1},  and a number of interesting
experimental and theoretical studies have been published subsequently; see,
for example, Glaberson \textit{et al} \cite{glaberson1} on an instability
in the presence of an axial flow of the normal fluid;  and the study of
non-linear effects,  leading to soliton behaviour \cite{hasimoto1}  and to
an associated sideband instability \cite{samuels1}.  Non-linear effects
remain interesting,  and an aspect of them that is important in our
understanding of Kelvin waves at very low temperatures,  and which may be of
rather general interest in non-linear dynamics, is discussed in this paper.

At temperatures where there is a significant fraction of normal fluid Kelvin
waves in superfluid $^4$He are damped by mutual friction, which is the
frictional force exerted on a vortex when it moves relative to the normal
fluid.  This paper is concerned with the expected behaviour of Kelvin waves
at  very low temperatures,  when damping due to mutual friction can be
neglected.  Under these conditions Kelvin waves can be damped only by
radiation of phonons,  but the damping is expected to be extremely small
\cite{vinen1} unless the frequency is very large,  typically of order 4 GHz
($k \sim$ 2 nm$^{-1}$).   Kelvin waves of lower frequency are essentially
undamped.  In these circumstances Kelvin waves of a particular low frequency
can lose energy only by non-linear coupling to waves of a different
frequency.  We are led therefore to consider the following situation.
Suppose that we have a rectilinear vortex of finite length,  and that we
continuously drive one low-frequency Kelvin mode on this vortex.  The mode
will grow in amplitude,  until non-linear effects give rise to a transfer of
energy to other modes,  particularly at higher frequencies.  This process
will presumably continue until modes are excited that have a frequency
sufficiently high for effective phonon radiation.  The aim of this paper is
to predict the details of this process,  by means largely of computer simula
tions.  We shall consider,  for example,  whether there is a steady state,
in which energy is injected at a low frequency and dissipated by phonon
emission at a high frequency,  and whether there is some well-defined and
simple spectrum of the Kelvin waves existing in this regime.  We shall
regard this regime as a cascade, although we are not sure whether it is a
strict cascade in the sense that energy is transferred \textit{in steps} to
higher wavenumbers.  We shall find that a well-defined spectrum does seem to
exist, that it is simple in form, and that it is remarkably insensitive to
the amplitude and frequency of  the drive.

This behaviour is interesting in its own right.  However, we ourselves were
led to investigate it by an interest in the decay of turbulence in
superfluid $^4$He at very low temperatures \cite{vinen2}, where mutual
friction has a negligible damping effect on vortex motion.  As in classical
turbulence, energy in superfluid turbulence must probably flow from larger
to smaller length scales,  and it has been suggested that on the smallest
scales the relevant motion is a Kelvin wave on a vortex with wavenumber
greater than the inverse vortex spacing.  It is then of interest to
understand how energy can flow in a system of Kelvin waves towards higher
wavenumbers,  where it can ultimately be dissipated at the highest
wavenumber by phonon radiation. In the context of superfluid turbulence the
details are likely to be quite complicated (Kelvin waves are likely to be
generated by vortex reconnections),  and it is not yet generally accepted
that a Kelvin-wave cascade is strictly necessary.  Nevertheless it seems
likely that such a cascade has a significant role.  These matters have been
discussed in a recent review \cite{vinen2},  and we plan further discussion
in forthcoming papers.

We believe that the behaviour we find may exist also in other types of
system and may therefore have other applications. It should be explained
that similar problems to that discussed here have been addressed by other
authors\cite{svistunov1,araki1,kivotides1,nemirovskii1}, but they relate to
situations that are significantly different;  for example, where there is no
steady state or where the length of vortex line is constrained not to
increase.  Thus Araki and Tsubota \textit{et al} \cite{araki1} carried out
numerical simulations on an initial configuration in which a vortex ring
approaches a rectilinear vortex,  Kelvin waves being generated by
reconnections when collision takes place.  There was no steady input of
energy and no obvious dissipation.  The numerical simulations of Kivotides
\textit{et al} \cite{kivotides1} related to an initial configuration of four
vortex rings,  Kelvin waves again being generated by reconnections when the
rings collide.  Again there was no steady input of energy and no obvious
dissipation. In both these simulations the authors found some evidence for
the development of a spectrum similar to Eq. (6) below,  although the
situation is obviously different from that considered in this paper.   The
computational work of Nemirovskii \textit{et al} \cite{nemirovskii1} did
deal with a steady state in which waves on a vortex ring were generated by
low-frequency noise, although the total length of line was kept artificially
constant;  the results suggested the existence of a spectrum similar to Eq.
(3) below,  although it did not exhibit the insensitivity to the driving
conditions that we ourselves find.  Generally, then,  it seems that this
other work has not led to such simple and clear-cut results as we report in
this paper,  and, as we shall discuss in later papers, it  may  be less
relevant to the decay of superfluid turbulence, especially when, as is often
likely to be the case \cite{vinen2},  energy flows from a large reservoir
associated with large-scale quasi-classical turbulent motion into motion on
a scale less than the vortex-line spacing.

 We consider a model system in which the helium is contained in the space
between two parallel sheets, separated by distance $\ell_{B}= 1$ cm,  with a
single,  initially rectilinear,  vortex is stretched between opposite points
on the two sheets.  Kelvin waves can be excited on this vortex,  and
periodic boundary conditions are applied at each end.  Thus the allowed
wavenumbers of the Kelvin waves are given by
%Eq.2
\begin{equation} k=\frac{2\pi n}{\ell_{B}},
\end{equation}
where $n$ is an integer ($>0$). We imagine that one of these modes, with $n$
equal to a small integer $n_{0}$, is continuously driven, so that its
amplitude tends continuously to increase. As the amplitude increases
non-linear coupling to other modes sets in,  and we can expect energy to
flow from the mode $n_{0}$ to other modes, with both larger and smaller
wavenumbers.  Now we introduce a suitably strong damping for all modes with
$n$ exceeding a large critical value $n_{c}$. This is intended to mimic the
effect of phonon emission,  although, because of inevitable computational
limitations, it is occurring at a much smaller frequency.  Then we ask
whether there is a steady state, described by an energy spectrum $E_{k}$, in
which the energy input to the mode $n_{0}$ is balanced by dissipation in the
modes with $n>n_{c}$. We find that such a steady state does seem to exist,
and we determine the character of the corresponding energy spectrum. We
observe no reconnections.

An important feature of the model lies in the fact that Kelvin modes with
wavenumbers less than $2\pi/\ell_{B}$ cannot be excited,  so that energy
cannot flow to smaller and smaller wavenumbers.  The relevance of this
feature to the decay of superfluid turbulence will be discussed in a later
paper.

The simulations are based on the vortex filament model,  and they are
similar to those described by Schwarz \cite{schwarz1} and used in more
recent work by one of the authors \cite{tsubota1}.  The undisplaced vortex
lies along the $z$-axis. Its calculated motion is based on the full
Biot-Savart law and therefore takes account of both local and non-local
contributions. The force that drives one mode is of the form
$V\rho\kappa\sin(k_{0}z-\omega_{0}t)$, where $k_{0}=2\pi n_{0}/\ell_{B}$,
$\rho$ is the density of the helium,  and $\omega_{0}$ is related to $k_{0}$
by the dispersion relation (1).   Damping at the highest wavenumber allowed
by the resolution of the simulations (1/60 cm) is applied by a periodic
smoothing process,  the details of which will be described in a later
publication;  this allows an effective dissipation at the highest wavenumber
that can adapt to the flux of energy through $k$-space arising from the
drive.

Fig. 1 shows how the total length of line evolves in time after application
of the driving force.  We see that it reaches a steady average value,
suggesting the existence of a steady state.
%Fig 1
\begin{figure}[btp]
\includegraphics[height=0.30\textheight]{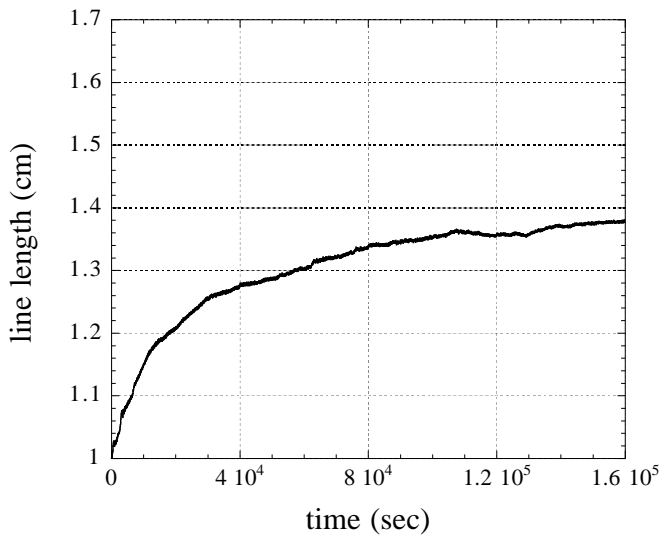}
\caption{The development in time of the total length of vortex line.}
\label{line length}
\end{figure}
%Fig 2
\begin{figure}[btp]
\includegraphics[height=0.30\textheight]{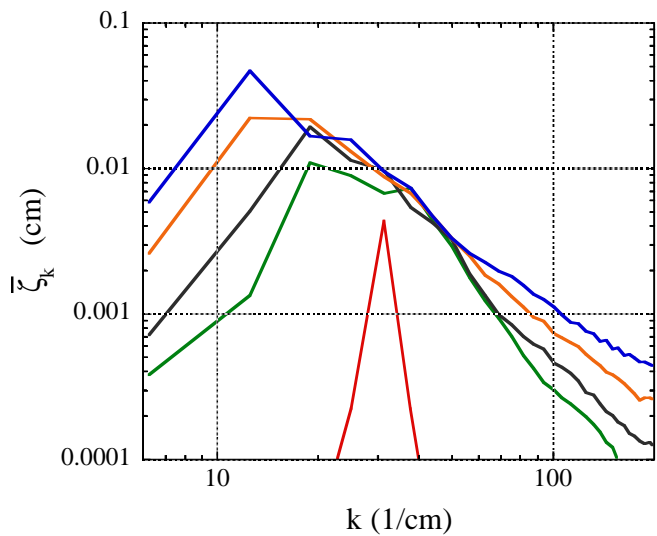}
\caption{Time development of $\bar{\zeta}_{k}(t)$.  The red, green, black,
orange and blue lines refer respectively to averages over 0-800 sec,
10000-10800 sec, 20000-20800 sec, 40000-40800 sec and 140000-140800 sec. The
blue line relates to the steady state.}
\label{time}
\end{figure}

We express our more detailed results in terms of the root mean square
amplitudes $\bar{\zeta}_{k}(t)=\langle\zeta^{*}_{k}\zeta_{k}\rangle^{1/2}$
of the Fourier components of the displacement of the vortex.  Figure 2 shows
how these amplitudes develop in time after the application of a drive with
$V=2.5\times10^{-5}$ cm s$^{-1}$ and $k_{0}=10\pi$ cm$^{-1}$.  We see that
initially only the mode that resonates with the drive is excited.  However,
as time passes,  non-linear interactions lead to the excitation of all
other modes.  Eventually the spectrum reaches a steady state,  shown by the
blue line,  and there is then no further change.   In this steady state
energy is injected at a certain rate at the wavenumber $k_{0}$, and it is
dissipated at the same rate at thehighest
wavenumber.  For large values of $k$,  where the modes form practically a
continuum,  the steady state is observed to have,  to a good approximation,
a spectrum  of the simple form
%Eq 3
\begin{equation} \bar{\zeta}_{k}^{2}=A\ell_{B}^{-1} k^{-3}.
\end{equation}
where the dimensionless parameter $A$ is of order unity.
%Fig 3
\begin{figure}[btp]
\includegraphics[height=0.30\textheight]{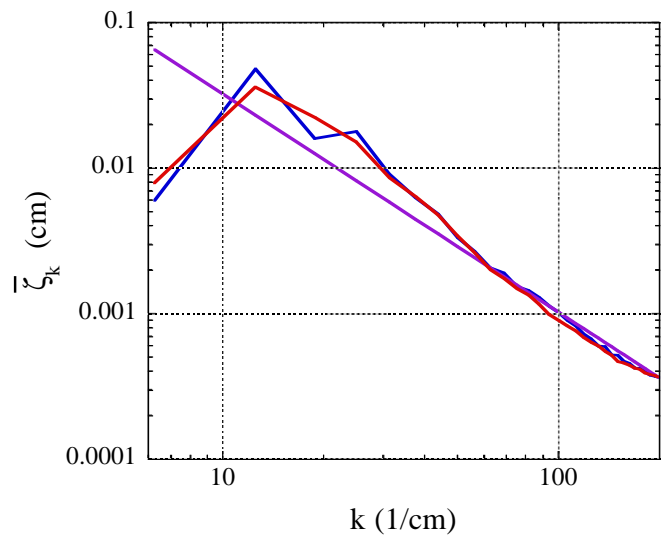}
\caption{Steady-state values of $\bar{\zeta}_{k}(t)$ for two different drive
amplitudes.  The blue and red lines are for,  respectively,
$V=2.5\times10^{-5}$ cm sec$^{-1}$ and $V=2.5\times10^{-4}$ cm sec$^{-1}$.
The mauve line has the form of Eq. (3).}
\label{Different drive}
\end{figure}
%Fig 4
\begin{figure}[btp]
\includegraphics[height=0.30\textheight]{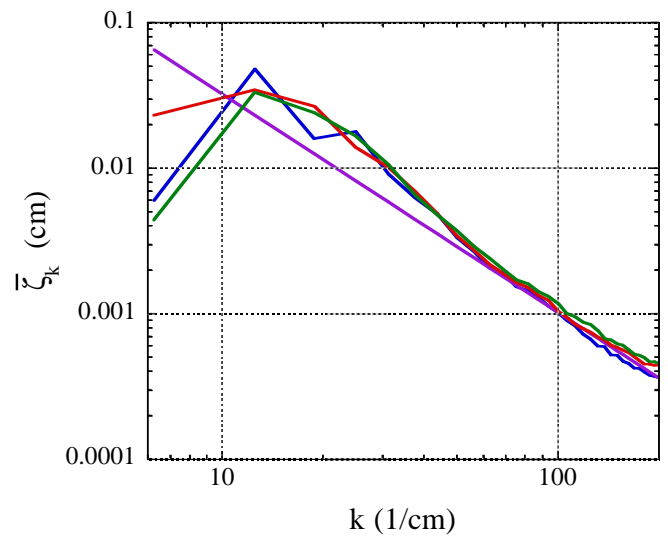}
\caption{Steady-state values of $\bar{\zeta}_{k}(t)$ for drives at three
different wavenumbers. The red, green and blue lines refer respectively to
$k_{0}=2\pi$ cm$^{-1}$, $k_{0}=4\pi$ cm$^{-1}$ and $k_{0}=10\pi$ cm$^{-1}$.
Again the mauve line has the form of Eq. (3).}
\label{Different wavenumber}
\end{figure}

Figs. 3 and 4  show the effects respectively of increasing the drive
amplitude $V$ by a factor of ten and of changing the drive wavenumber
$k_{0}$.  We see that there is no effect on the steady state,  within the
error of the simulations,  at least at the higher wavenumbers.  The steady
state takes longer to be established at the lower drive amplitude, which
suggests that even with a small drive amplitude the same steady state would
be established after a sufficiently large time;  but limitations on the time
available for a simulation have not allowed us to check this suggestion.

Important questions relate to the relationship between the drive amplitude
and the power input to the system of Kelvin waves.  For small times after
the drive is first established non-linearities in the system are relatively
unimportant,  and the power input can be calculated from the product of the
drive amplitude and the amplitude of the velocity response at wavenumber
$k_{0}$, taking proper account of phase differences.  For later times,
however, the system exhibits very strong non-linear behaviour,  and this
simple technique is no longer applicable.  Energy seems to be injected
through processes involving two wavevectors,  the difference between which
is equal to $k_{0}$.  Therefore we have not yet been able to calculate how
the power input varies with the drive amplitude, although we hope to do so
later.  Careful analysis of the operation of damping in our model for large
wavenumbers might allow us to obtain the rate of dissipation of energy,  but
we have not yet completed this analysis.  For the present we can only make
the reasonable assumption that an increase in the drive amplitude does
increase the power input.   We can conclude with some confidence therefore
that the spectrum is insensitive to the drive amplitude, the drive
frequency, and the power input from the drive.

The mean energy per unit length of vortex in a mode $k$ is related to
$\bar{\zeta}_{k}$ by the equation
%Eq.4
\begin{equation}  E_{k} = \epsilon_{K} k^{2} \bar{\zeta}_{k}^{2}
\end{equation}
where $\epsilon_{K}$ is an effective energy per unit length of vortex,
given by
%Eq.5
\begin{equation} \epsilon_{K} = \frac{\rho \kappa^{2}}{4 \pi}
\bigg[\ln\bigg(\frac{1}{ka}\bigg)+c_{1} \bigg].
\end{equation}
It follows from Eqs. (3) and (4) that
%Eq.6
\begin{equation}  E_{k} =A\epsilon_{K}(k\ell_{B})^{-1}.
\end{equation}

We conclude that in our model system a steady state cascade does develop,
that this state is characterized by the energy spectrum (6),  and that,
remarkably,  this spectrum is insensitive to the frequency and amplitude of
the drive and to the power input at the drive frequency.

It is interesting to ask what physics underlies this result.    We suggest
that, in the steady state and for waves of wavenumber of order $k$,  there
is a saturation in the local amplitude of the Kelvin waves at a value of
roughly $k^{-1}$;  this arises from a sudden onset of strong non-linear
effects when the amplitude is of order the wavelength.  We make the
reasonable assumption that the total mean square amplitude of the
displacement of the vortex is independent of the length ($\ell_{B}$) of the
vortex. The spectrum (3) must then be proportional to $\ell_{B}^{-1}$.  Our
assumption about the sudden onset of non-linear effects means that the only
other parameter on which this spectrum can depend is $k$.  The form (3) then
follows from a dimensional argument.
We propose now to investigate whether this type of behaviour can be found in
other forms of wave propagation,  at least if they have the same types of
dispersive and non-linear characteristics (which are known to lead to
soliton behaviour \cite{hasimoto1}).

As far as we are aware,  there are as yet no experimental results that
relate directly to the behaviour of Kelvin waves in superfluid $^4$He at
very low temperatures.  Such experiments,  which might involve the
excitation of Kelvin waves on the regular array of vortices existing in the
uniformly rotating liquid,  as in the early experiments of Hall
\cite{hall1}, would be of great interest.  Measurements might be made of the
rate at which energy is transmitted to the array from a suitable transducer,
and of the rate at which energy eventually appears as heat after the
Kelvin-wave cascade has been established.

In summary we have reported the results of computer simulations of the
behaviour of Kelvin waves on a rectilinear quantized vortex of finite length
in superfluid $^4$He at a temperature so low that the waves suffer no
attenuation from mutual friction with the normal fluid,  the only
attenuation arising from phonon radiation at a very high frequency.  The
waves are excited by continuousy driving the system at a small wavenumber.
The amplitude of the driven mode increases until non-linear coupling leads
to a transfer of energy to all other modes.  A steady state is established,
described by a simple energy spectrum,   the form of which is remarkably
insensitive to the strength and other details of the drive.

\begin{acknowledgements}
We are grateful for stimulating conversations and correspondence on the
general background to this work with Carlo Barenghi, Demos Kivotides, Boris
Svistunov and Sergey Nemirovskii.  MT is grateful to the Japan Society for
the Promotion of Science for financial support through the Japan-UK
Scientific Cooperative program (Joint Research project)  and through a
Grant-in-Aid for Scientific Research (Grant no. 15340122).  WFV is grateful
for financial support through a Royal Society Joint Project on Experimental
and Theoretical Study of Superfluid Turbulence.
\end{acknowledgements}

\end{document}